\documentclass[structabstract]{aa}  

\usepackage{graphicx}
\usepackage{txfonts}
\usepackage{indentfirst}
\usepackage{natbib}
\usepackage{ulem}
\bibpunct{(}{)}{;}{a}{}{,} 
\usepackage[scriptsize]{subfigure}
\usepackage{amsmath1}
\usepackage{multirow}

\def\ltsima{$\; \buildrel < \over \sim \;$}
\def\simlt{\lower.5ex\hbox{\ltsima}}      

\def\gtsima{$\; \buildrel > \over \sim \;$}
\def\simgt{\lower.5ex\hbox{\gtsima}}

\usepackage{color}
 
\begin{document}

  \title{Synthetic observations of first hydrostatic cores in collapsing low-mass dense cores.}
\subtitle{II. Simulated ALMA dust emission maps}  
   \author{B. Commer\c con
           \inst{1},
          F. Levrier 
          \inst{1},
          A.J. Maury
          \inst{2},
         Th. Henning
	 \inst{3}, and
	R. Launhardt
         \inst{3}
           }
   \offprints{B. Commer\c con\\{\tt benoit.commercon@lra.ens.fr}}

\institute{Laboratoire de radioastronomie, UMR 8112 du CNRS, \'{E}cole Normale Sup\'{e}rieure et Observatoire de Paris, 24 rue Lhomond, 75231 Paris Cedex 05, France
                        \and
ESO, Karl Schwarzschild Strasse 2, 85748, Garching bei M\"unchen, Germany
                        \and
                        Max-Planck-Institut f{\"u}r Astronomie, K{\"o}nigstuhl 17, 69117 Heidelberg, Germany}

   \date{Received 20 july 2012; accepted 3 october 2012}

  \abstract  
  {First hydrostatic cores are predicted by theories of star formation,  but their existence has never been demonstrated convincingly by (sub)millimeter observations. Furthermore, the multiplicity at the early phases of the star formation process is poorly constrained.}   
  {The purpose of this paper is twofold. First, we seek to provide predictions of ALMA dust continuum emission maps from early Class 0 objects. Second, we show to what extent ALMA will be able to probe the fragmentation scale in these objects.}
  {Following our previous paper (Commer\c con et al. 2012, hereafter paper I), we post-process three  state-of-the-art radiation-magneto-hydrodynamic 3D adaptive mesh refinement calculations to compute the emanating dust emission maps. We then produce synthetic ALMA observations of the dust thermal continuum from first hydrostatic cores.}
      {We present the first synthetic ALMA observations of dust continuum emission from first hydrostatic cores. We analyze the results given by the different bands and configurations and we discuss for which combinations of the two the first hydrostatic cores would most likely be observed. We also show that observing dust continuum emission with ALMA will help in identifying the physical processes occurring within collapsing dense cores. If the magnetic field is playing a role, the emission pattern will show evidence of a pseudo-disk and even of a magnetically driven outflow, which pure hydrodynamical calculations cannot reproduce.}  
       {The capabilities of ALMA will enable us to make significant progress towards understanding fragmentation at the early Class 0 stage and discovering first hydrostatic cores.}   


\titlerunning{}
\authorrunning{B. Commer\c con et al.}
   \maketitle


\section{Introduction}

It is established that most stars form in multiple systems \citep{Duquennoy_Mayor_1991,Janson_et_al_2012}. This indicates a fragmentation process during star formation, which can be explained by several mechanisms \citep[e.g.,][]{Bodenheimer_et_al_2000,McKee_Ostriker_2007}. The first picture is to consider the interplay between turbulence and gravity within molecular clouds, which can lead to an initial fragmentation prior to the collapse \citep{Hennebelle_Chabrier_2008}. In this picture, the stellar Initial Mass Function (IMF) is mainly determined at the dense core formation stage, the latter undergoing collapse without fragmenting into individual objects \citep[e.g.,][]{Price_Bate_2007,Hennebelle_Teyssier_2008,Commercon_et_al_2010}. On the other hand, fragmentation may also occur during the collapse of molecular clouds \citep[e.g][]{Bate_Bonnell_2005,Bate_2012} or within disks that are formed because of the conservation of angular momentum \citep[e.g.,][]{Whitworth_2006,Commercon_et_al_2008}. The fragmentation process thus remains a matter of intense debate, and in particular, the disk formation and early fragmentation (i.e., during the early phase of the collapse) issues appear to be critical to better constrain the star formation mechanism \citep[e.g.,][]{Li_et_al_2011,Joos_et_al_2012,Seifried_et_al_2012}.

The tremendous combined developments of observational and supercomputing capabilities allow to study astrophysical processes on scales which were until today unresolved. In particular, great advances in understanding the star formation process have been achieved during the past ten years. On the one hand, thanks to various (sub)millimeter interferometric facilities (e.g., the IRAM Plateau de Bure Interferometer, PdBI, and the Submillimeter Array, SMA) and to the {\it{Spitzer }}\rm and {\it{Herschel }}\rm space telescopes, much progress has been achieved towards understanding the formation and structure of prestellar dense cores, and constraining the evolutionary stages of star-forming regions \citep[e.g.,][]{Kennicutt_Evans_2012}. On the other hand, numerical models of star formation integrate more and more physical processes. Among the most important ones, magnetic fields and radiative transfer appear to shape the collapse and fragmentation of prestellar dense cores \citep[e.g.,][]{Hennebelle_Teyssier_2008, Bate_2009}, while their combined feedback dramatically inhibits fragmentation in low- and high-mass collapsing dense cores \citep{Commercon_et_al_2010,Commercon_et_al_2011c}. Unfortunately, there is currently no direct evidence of these mechanisms, since observations are not yet able to probe the fragmentation scale in nearby star-forming regions. 
While wide $\simgt$5000 AU multiple systems are often detected around the youngest (Class 0) protostars \citep[e.g.,][]{Chen_et_al_2008,Launhardt_et_al_2010}, 
the highest resolution observations so far with synthesized half power beamwidths of $0.3" -1"$ (e.g., using the IRAM PdBI or the SMA) 
show a lack of close $\simlt$2000~AU multiple systems \citep{Maury_et_al_2010}. However, at a distance of the nearest star-forming regions (e.g., 140 pc for Taurus), 
this angular resolution probes only linear scales larger than 40-50 AU. In order to probe smaller scales, Atacama Large Millimeter/Submillimeter Array 
(ALMA) observations are definitely needed. 

First hydrostatic cores (FHSC), i.e., first Larson cores \citep{Larson_1969} are the first protostellar objects formed during the star formation process with typical sizes of a few AU. Although their existence is predicted by theory \citep[e.g.,][]{Larson_1969,Masunaga_et_al_1998,Tomida_2010,Commercon_et_al_2011b}, there still is no strong observational evidence for such objects, because FHSCs are deeply embedded within collapsing cores and their lifetimes are relatively short \citep[at most a few thousand years, e.g.,][hereafter Paper I]{Tomida_2010b,Commercon_et_al_2012} compared to the Class 0 phase duration \citep[$0.1-0.2$ Myr,][]{Evans_et_al_2009}. Several candidate FHSCs are known \citep[e.g.,][]{Belloche_et_al_2006,Chen_et_al_2010,Pineda_et_al_2011,Chen_et_al_2012}, but none has yet been confirmed.   
We showed in Paper I that spectral energy distributions (SEDs) of star-forming clumps can help in identifying FHSC candidates, but are not able to assess the physical conditions within the cores and in particular the level of fragmentation. For that, high resolution interferometric imaging is needed. It is also unclear to what extent FHSCs can be distinguished from more evolved very-low luminosity objects \citep[VeLLOs;][]{DiFrancesco_et_al_2007} and second hydrostatic cores (i.e., the prostostars).

\begin{table*}[t]
\caption{Summary of the physical structures and FHSC lifetimes found in the three models. The last four columns indicate whether these structures are observed (cross) or not (dash) in the synthetic ALMA observations presented in Fig. \ref{c15_b3} to \ref{c20_b4}.}
\label{table:summary}
\centering
\begin{tabular}{cccccccc}
\hline\hline          
\multirow{2}{*}{Model}   & \multirow{2}{*}{Physical structures} & \multirow{2}{*}{FHSC lifetime} & \multicolumn{4}{c}{Structures in synthetic observations} \\
   & & & $\theta=0^\circ$  & $\theta=45^\circ$ &$\theta=60^\circ$ &$\theta=90^\circ$ \\
\hline
\multirow{3}{*}{MU2} & one object      &\multirow{3}{*}{1.2 kyr} & x & x & x &x                                                     \\
				 & pseudo-disk   &  & x & x  & x &x                                                     \\
				 & outflow            & &- & - & - & -                                                     \\
\hline
\multirow{4}{*}{MU10} & one object &\multirow{4}{*}{1.8 kyr} &  x & x & x &x                                                     \\
				 & pseudo-disk  & & x & x & x &x                                                     \\ 
				 & compact disk                  & & x & x & x & -                                                     \\
				 & outflow            & & - & - & - &x                                                     \\
\hline    
\multirow{2}{*}{MU200} & five fragments  & \multirow{2}{*}{$> 3$ kyr} & x & x & x & -                                       \\
				 & disk                          &  & x & x & x &x                                       \\
\hline    
\end{tabular}
\end{table*}

To address this necessity   of theoretical predictions for the appearance of early phases of star formation, \cite{Krumholz_2007b}, \cite{Semenov_2008}, \cite{Cossins_2010}, and \cite{Offner_2012} presented synthetic ALMA observations of dust continuum or line emission. In this paper, we present the first predictive dust emission maps of embedded FHSCs as they should be observable with ALMA. This study focusing on dust continuum is considered as a first step towards FHSC characterization in combination with the results of Paper I.

The paper is organized as follows. Section 2 presents the physical models and the method we use to derive synthetic ALMA dust emission maps. Section 3 reports on the results we obtain to select the best ALMA configuration and the best receiver band in order to observe FHSCs. We discuss the limitation of our work in Sec. 4. Section 5 presents our conclusions and perspectives concerning the future work needed to confirm the results obtained as a first step with dust continuum observations.

\section{Method}

In this study, we restrict our work to the early stages of the star formation process, i.e., the first collapse and FHSC formation. As mentioned in the introduction, a lot of progress has been achieved in theory and observations to characterize what we would expect at those early stages. In the following, we combine state-of-the-art tools to produce synthetic ALMA dust emission observations.

\subsection{The physical models}

We performed 3D full radiation-magneto-hydrodynamic (RMHD) calculations using the adaptive mesh refinement code {\ttfamily{RAMSES }}\rm \citep{Teyssier_2002}, which integrates the equations of ideal magneto-hydrodynamics \citep{Fromang_et_al_2006} and uses the grey flux-limited-diffusion approximation for the radiative transfer \citep{Commercon_et_al_2011a}. We used the same RMHD calculations presented in Paper I, which consisted in letting rotating (in solid-body rotation) 1 M$_\odot$ dense cores collapse, with initially uniform temperature, density and magnetic field. The ratio of the initial thermal and rotational energies to the gravitational energy are respectively $\alpha=0.35$ and $\beta=0.045$. We ran three different models with the same initial conditions except for the initial magnetization, which is parametrized by the mass-to-flux to critical mass-to-flux ratio $\mu=(M_0/\Phi)/(M_0/\Phi)_{\rm c}$. The three models are depicted as follows :  MU2 model (strong magnetic field, $\mu=2$), MU10 model (intermediate magnetic field, $\mu=10$) and MU200 model (quasi-hydro case, $\mu=200$). These three models are representative of the diversity of FHSCs and of their environments (disk, pseudo-disk, and outflow) that are predicted by the theory. The different physical structures and FHSC lifetimes found in the three models are summarized in Table \ref{table:summary}. In the MU2 model, only one FHSC is formed, surrounded by a pseudo-disk, and an outflow has been launched. The MU10 model does not fragment either and results in a system composed of a disk, a pseudo-disk and an outflow. The MU200 model has classical features of hydrodynamical models, in which relatively large disks ($\sim 150$ AU) are formed and subsequently fragment. Readers are referred to Paper I for a thorough description of the different models and their limitations. Note that the MU2 and MU10 models are more representative of the observed magnetization level in star forming regions \citep[i.e., $\mu\approx2-3$,][]{Falgarone_et_al_2008,Crutcher_et_al_2010}, even though determining magnetic field strength remains challenging.

In the following, the calculations are post-processed at the same times as in Fig. 2 of Paper I. These are representative of the characteristics of the three models, i.e., $t_0+0.78$ kyr for the MU2 model, $t_0+1.81$ kyr for the MU10 model, and $t_0+3.26$ kyr for the MU200 model (where $t_0$ corresponds to the FHSCs formation in each model, i.e. 50.4 kyr for MU2, 35.7 kyr for MU10 , and 35.4 kyr for MU200).

\subsection{Dust emission models with {\tt RAMDC-3D}}

We used the same interface as that presented in Paper I, which couples the outputs of the RMHD calculations, done within {\tt RAMSES}, to the 3D radiative transfer code {\tt RADMC-3D}\footnote{\tt \tiny www.ita.uni-heidelberg.de/$\sim$dullemond/software/radmc-3d/}. In this interface, we assumed that gas and dust are thermally coupled $(T_\mathrm{dust}=T_\mathrm{gas})$, which is a valid approximation given the high density within the dense cores \citep[e.g,][]{Galli_et_al_2002}.  We also assumed that the gas temperature $T_\mathrm{gas}$ computed in the RMHD calculations is correct \citep{Commercon_et_al_2011b,Vaytet_et_al_2012}. We used the low temperature opacities of \cite{Semenov_et_al_2003A&A} for a model in which dust is made of homogeneous spheres with a "normal" ($\mathrm{Fe}/\mathrm{Fe+Mg}=0.3$) silicate composition.

The dust thermal continuum emission maps were computed on a square box with physical size $\Delta x=1300~\mathrm{AU}$ and a resolution $\delta x=2.54~\mathrm{AU}$. Since models were assumed to be at a distance $D=150~\mathrm{pc}$, this translates into an angular extent  $\Delta \vartheta=8.67"$ and an angular resolution $\delta \vartheta=0.017"$. Four different viewing angles are used in the following : $\theta=0^{\circ}$ (equatorial plane seen face-on), $\theta=45^{\circ}$, $\theta=60^{\circ}$, and $\theta=90^{\circ}$ (edge-on view, perpendicular to the rotational axis). These model dust emission maps were computed for six ALMA bands (3, 4, 6, 7, 8 and 9), including the four that are in use in the Early Science stage (see table~\ref{table:ALMA-bands}), over a total bandwidth $\Delta\nu=8~\mathrm{GHz}$ centered on the band's central frequency $\nu_0$. This was done by computing a set of ten emission maps, one for each of ten 800~MHz sub-bands, and averaging them. The resulting map gives the mean specific intensity $\left<I_\nu\right>$ of the thermal dust emission over the band, in erg cm$^{-2}$ s$^{-1}$ Hz$^{-1}$ sr$^{-1}$, and is then converted to a brightness temperature map in K via\footnote{The factor $10^{-3}$ comes from the CGS to SI conversion of the specific intensity, as the ALMA simulator assumes input brightness distributions to be in K.} 
$$
T_\mathrm{B}=10^{-3}\times\frac{c^2\left<I_\nu\right>}{2k_B\nu_0^2}.
$$

\begin{table}
\caption{Characteristics of the ALMA bands used, denoted $B$. Listed are the central frequency $\nu_0$, full bandwith $\Delta\nu_0$ (not to be confused with the bandwidth $\Delta\nu=8~\mathrm{GHz}$ accessible for a given observation), the field-of-view at the band center, and the availability of the band at the Early Science stage.}
\label{table:ALMA-bands}
\centering
\begin{tabular}{ccccc}
\hline\hline          
$B$ & $\nu_0$ [GHz] & $\Delta\nu_0$ [GHz] & FoV ["] & Early Science\\
\hline
3 & 100 & 32 & 63 & Yes\\
4 & 144 & 38 & 44 & No\\
6 & 243 & 64 & 26 & Yes\\
7 & 324 & 98 & 19 & Yes\\
8 & 442.5 & 115 & 14 & No\\
9 & 661 & 118 & 9.5 & Yes\\
\hline    
\end{tabular}
\end{table}


\begin{table}
\caption{Characteristics of the full ALMA configurations used, denoted $C$. Listed are the minimum and maximum baselines, and synthesized beam major and minor axes for bands 3 and 9.}
\label{table:ALMA-configs}
\centering
\begin{tabular}{ccccc}
\hline\hline          
$C$ & $b_\mathrm{min}$ [m] & $b_\mathrm{maj}$ [m] & $\theta_3$ ["] & $\theta_9$ ["] \\
\hline
5 & 15 & 390 & $2.47\times 2.12$ & $0.37\times 0.32$  \\
10 & 20 & 924 & $1.07\times 0.99$ & $0.16\times 0.15$  \\
15 & 25 & 1814 & $0.48\times 0.43$ & $0.07\times 0.06$  \\
20 & 49 & 3699 & $0.24\times 0.22$ & $0.04\times 0.03$  \\
\hline    
\end{tabular}
\end{table}

\begin{figure*}[htb]
\centering
\includegraphics[scale=0.4]{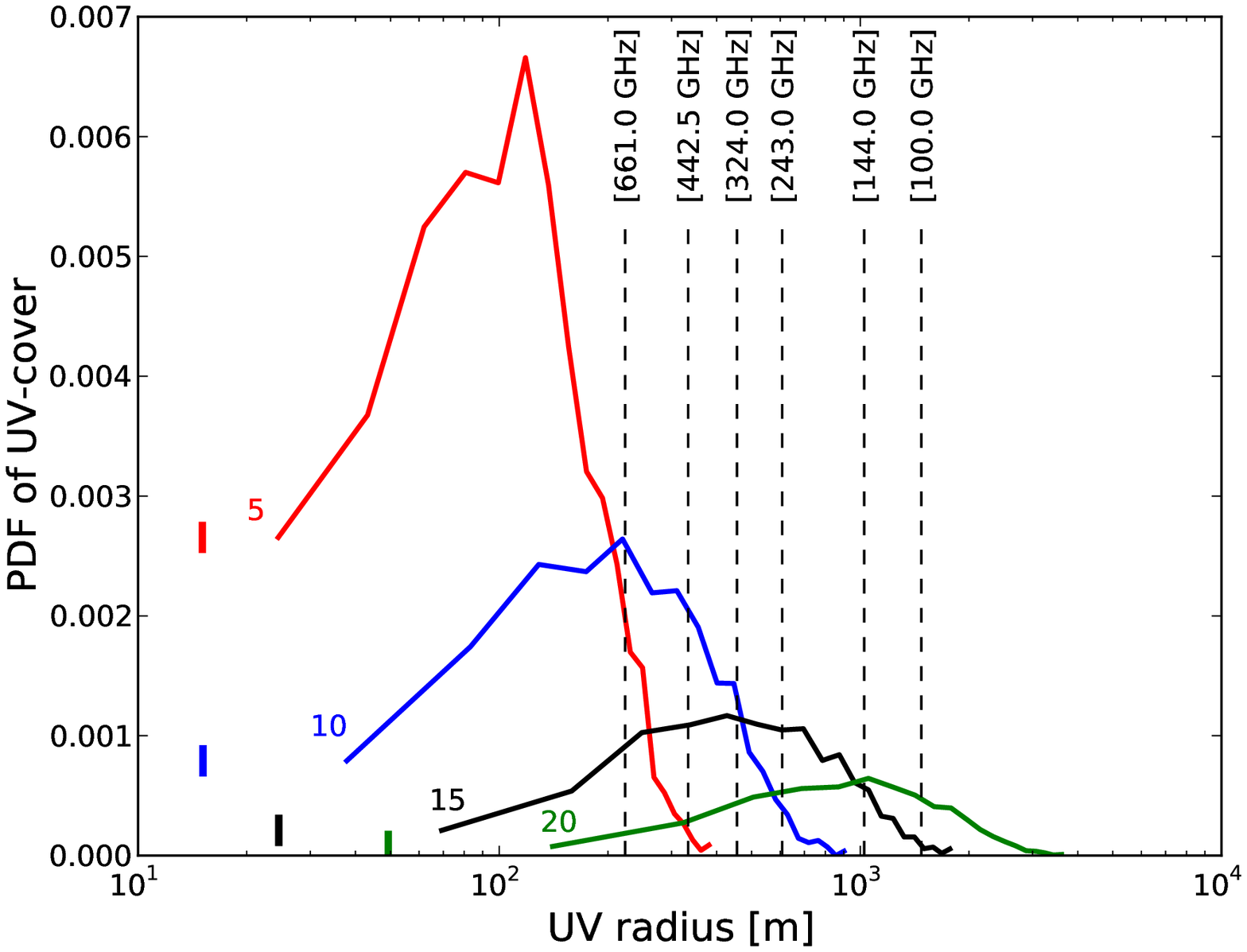}
\includegraphics[scale=0.4]{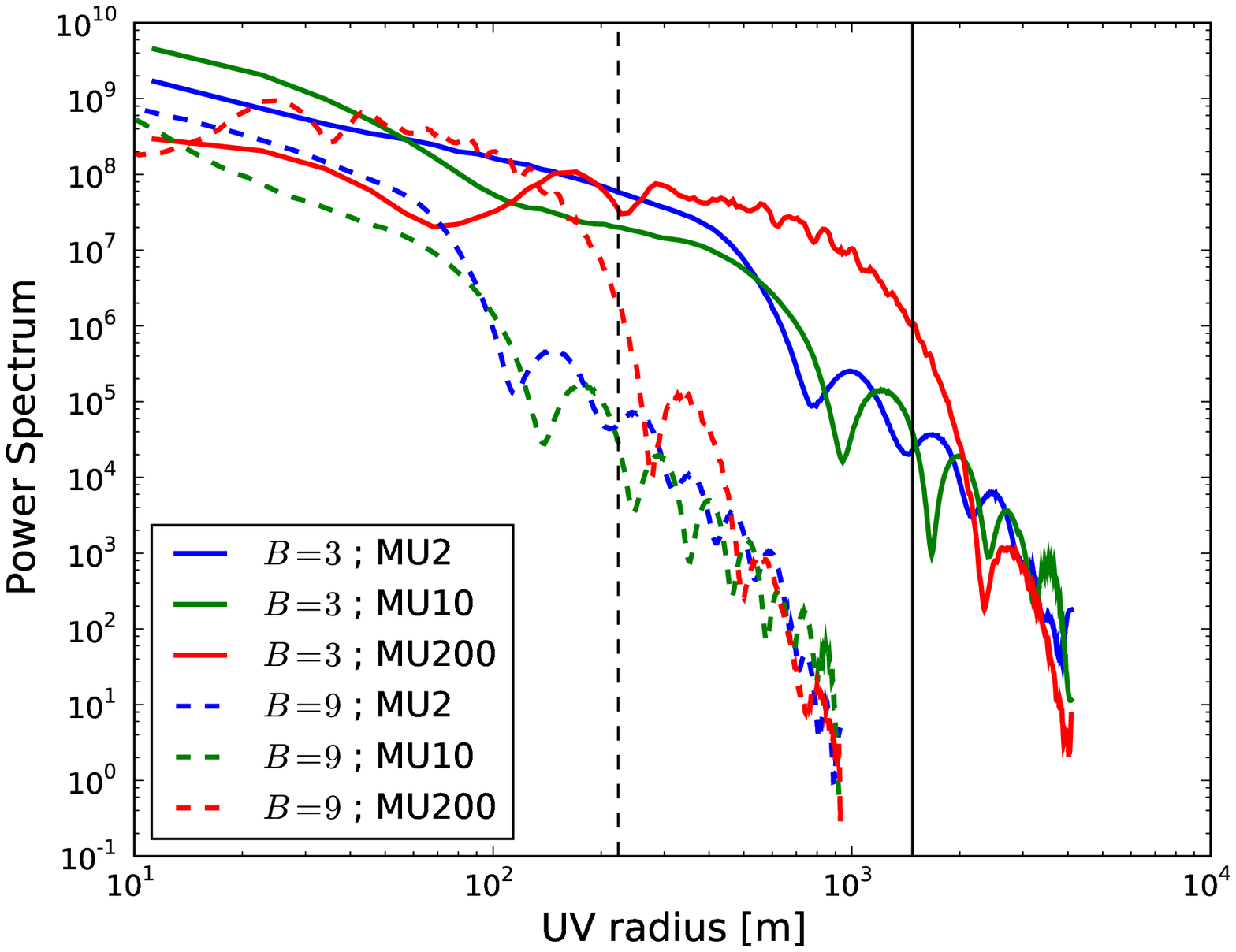}
\caption{{\it Left :} Probability density function from the distribution of samples in the $u-v$ plane for the 4 configurations used here (numbers 5, 10, 15 and 20, indicated next to each curve) as a function of $u-v$ radius $d_{uv}=\sqrt{u^2+v^2}$. The relevant parameters are the source position and duration of observation, which are described in section \ref{sec:method}. The small vertical lines indicate the minimum baseline for each configuration. Also marked, with dashed lines, are characteristic baselines $d_{uv}=cD/(2\pi\nu_0 d)$ corresponding to a physical size $d=10~\mathrm{AU}$ at a distance $D=150~\mathrm{pc}$, for the six frequencies $\nu_0$. {\it Right :} Power spectra of the input maps at 100 GHz (solid lines) and 661 GHz (dashed lines) for $\theta=0^\circ$. The wavenumber axis has been rescaled to match the $u-v$ radius of the left plot. The vertical black lines mark the positions of the 10 AU scale at 150 pc at these frequencies (solid for 100 GHz, dashed for 661 GHz, reported from left plot).}
\label{uvdistrib}
\end{figure*}

\begin{figure*}[t]
\includegraphics[scale=0.36]{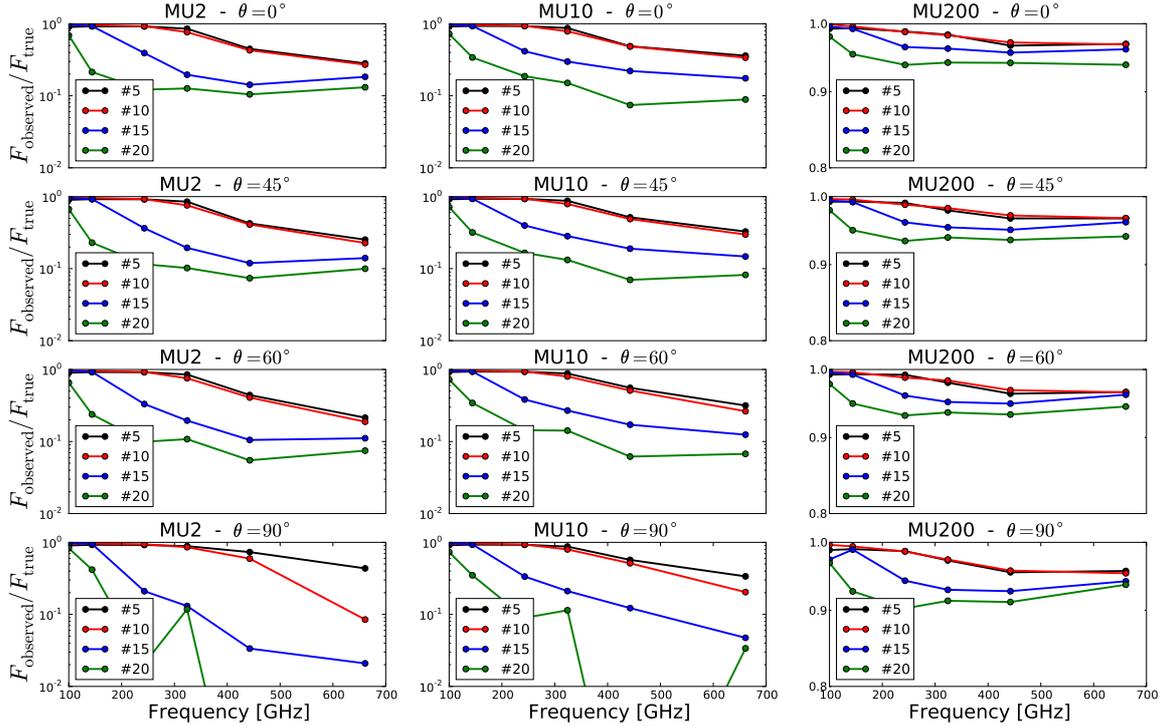}
\caption{Ratio $F_\mathrm{obs}/F_\mathrm{model}$ of the observed flux to the model flux as a function of frequency, for the three different models (MU2, MU10, and MU200), four array configurations ($C=5,10,15,20$), and four inclinations ($\theta=0^\circ, 45^\circ, 60^\circ, 90^\circ$). Magnetization level $\mu$ increases from left to right, and inclination angle $\theta$ increases from top to bottom. }
\label{fluxloss}
\end{figure*}

\subsection{Producing synthetic ALMA observations}
\label{sec:method}
The {\tt GILDAS} software package\footnote{\tt \tiny iram.fr/IRAMFR/GILDAS/}, developed and maintained by IRAM, is primarily intended for the reduction and analysis of observational data acquired via the IRAM instruments (single-dish 30-m radiotelescope at Pico Veleta and 6-antenna array at  IRAM PdBI). It also includes a full ALMA simulator, with up-to-date array configurations\footnote{We used release {\tt apr11h} of {\tt GILDAS}. The latest releases also contain Early Science Cycle 1 configurations of the array.}, which was primarily developed to assess the impact of the ALMA compact array (ACA) on the imaging capabilities of ALMA \citep{alma_memo_398,alma_memo_488}.



To produce synthetic ALMA observations from our dust emission models, maps are converted to the {\tt GILDAS} data format (GDF) and projected on a fixed sky position $(\alpha_0,\delta_0)=(0^\circ,-23^\circ)$ such that sources transit at  the zenith of the array. The field-of-view in each band (see table~\ref{table:ALMA-bands}), given by $1.22\lambda_0/d$ with $d=12~\mathrm{m}$ the antenna diameter and $\lambda_0$ the central wavelength of the band, shows that a single pointing suffices to map the emission in all cases. 

ALMA consists of two sub-arrays, the fifty 12-m antennas and the twelve 7-m antennas of the ACA. The array operation will consist in constantly moving antennas around, so that no two observations may be obtained with exactly the same configuration. We do not consider the compact array and focus on imaging the cores' thermal emission with ALMA only. To keep the number of simulations down to a manageable number, we select four "typical" configurations (out of the 28 representative configurations implemented in GILDAS), whose properties are listed in table~\ref{table:ALMA-configs}. Since we make one simulation per model, per configuration, per inclination angle, and per band, the total number of simulations is thus $3\times 4\times 4 \times 5 =240$. In all cases, model sources are observed for a total integration time of 18 minutes centered on the transit.

The ALMA simulator is a full pipeline, including the deconvolution of dirty images via a CLEAN algorithm \citep{clark80}. Outputs can thus be compared directly to input images smoothed to the same resolution, which the simulator also provides.

\section{Results}

\subsection{A necessary compromise}

In our models, thermal dust emission at millimeter and submillimeter wavelengths mostly comes from structures that are larger than the FHSC (i.e., from the disk and the pseudo-disk). Yet, to discriminate between different magnetization levels $\mu$, one needs to resolve the fragmentation scale of a few AU. This latter constraint means that only extended configurations of the array will be able to provide a sufficient angular resolution, in the few tens of milliarcseconds range. However, this comes with an increase of the minimum baseline length, so that large-scale emission is lost due to the central hole in the visibility (Fourier) space, also called $u-v$ plane.

To make this  idea more quantitative, Fig.~\ref{uvdistrib} ({\it left}) shows the distribution of visibility samples as a function of radius in the $u-v$ plane for the four configurations.  Marked on this graph are the baselines $d_{uv}=cD/(2\pi\nu_0 d)$ corresponding to a physical size $d=10~\mathrm{AU}$ (roughly equal to the size of fragments in the MU200 models) at a distance $D=150~\mathrm{pc}$, for the six central frequencies $\nu_0$. On the right plot of Fig.~\ref{uvdistrib} we display the power spectra of the face-on input maps at 100 GHz (band 3) and 661 GHz (band 9), with the wavenumber axis $k$ rescaled to match that of the $u-v$ radius on the left plot. This is done by noticing that the largest wavenumber in the power spectrum corresponds to the pixel physical size $\delta x$. 

 The fragmentation in the MU200 model appears as the excess power at intermediate scales compared to MU2 and MU10. This excess is clearly seen for $300~\mathrm{m}\lesssim d_{uv}\lesssim 2000~\mathrm{m}$ at 100 GHz, with $d_{uv}= 2000~\mathrm{m}$ roughly corresponding to the 10 AU scale at 150 pc. As far as the four configurations used here are concerned, this range of baselines is beyond $C=5$ and best probed by configurations $C=15$ and $C=20$. However, both of these configurations have a larger central hole than $C=5$ and $C=10$, and therefore should lose a larger amount of flux. This flux loss becomes larger at higher frequencies, since $d_{uv}\propto 1/\nu_0$ implies a global compression of the sources' power spectra towards smaller baselines.

For a global view on the emission loss at large scales, Fig.~\ref{fluxloss} shows the ratio between the observed flux to the model flux as a function of frequency and array configuration. As expected, the most compact configurations ($C=5$ and $C=10$) allow to recover essentially all of the flux in bands 3 to 7. Only in band 8 and 9 we see a $\sim 50\%$ drop in the received flux for the non-fragmenting models MU2 and MU10. Figure \ref{fluxloss} also confirms that flux loss increases with frequency in configurations 15 and 20, most ($\sim 80\%$) of the flux being lost already at 144 GHz in configuration 20. 
 Overall, it appears that observing in bands 3 and 4 with configuration 15 may provide the best compromise, as the flux loss is very limited (less than 8\%) and according to the left plot of Fig.~\ref{uvdistrib}, it should be able to resolve the fragmentation scale, at least in band 4.

\begin{figure*}
\includegraphics[scale=0.52]{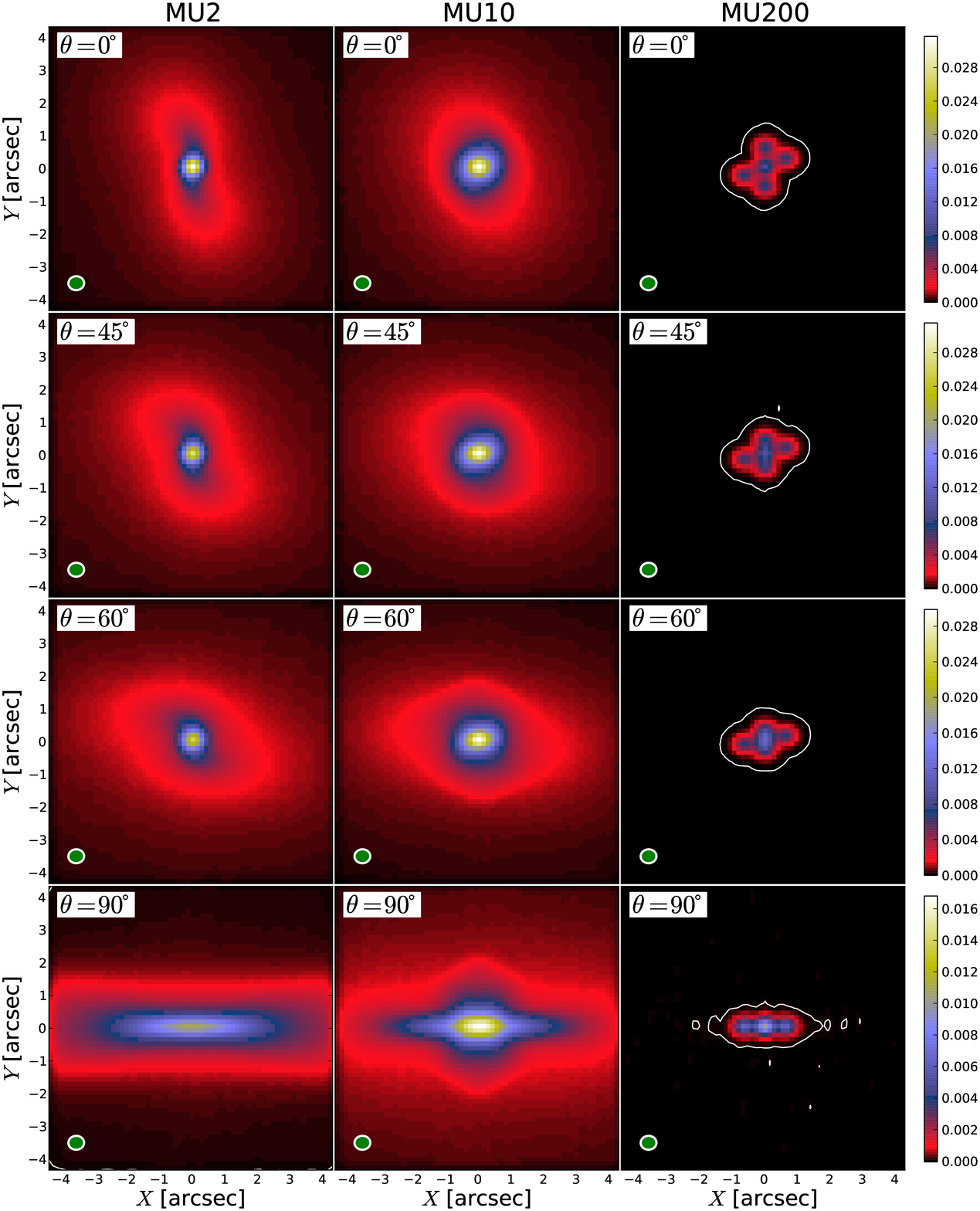}
\caption{Brightness distribution maps, in Jy/beam, output by the ALMA simulator in band $B=3$ (100 GHz) and array configuration $C=15$. Magnetization level $\mu$ increases from left to right, and inclination angle $\theta$ increases from top to bottom. Contours show the $3\sigma_S$ sensitivity limit in this band, as given by the ALMA Sensitivity Calculator. Here $\sigma_S=14.55~\mu{\mathrm{Jy}}$. The synthesized beam is shown in the bottom left corner of each plot. For a given $\theta$, color scales are identical across all columns.}
\label{c15_b3}
\end{figure*}
\begin{figure*}
\centering
\includegraphics[scale=0.52]{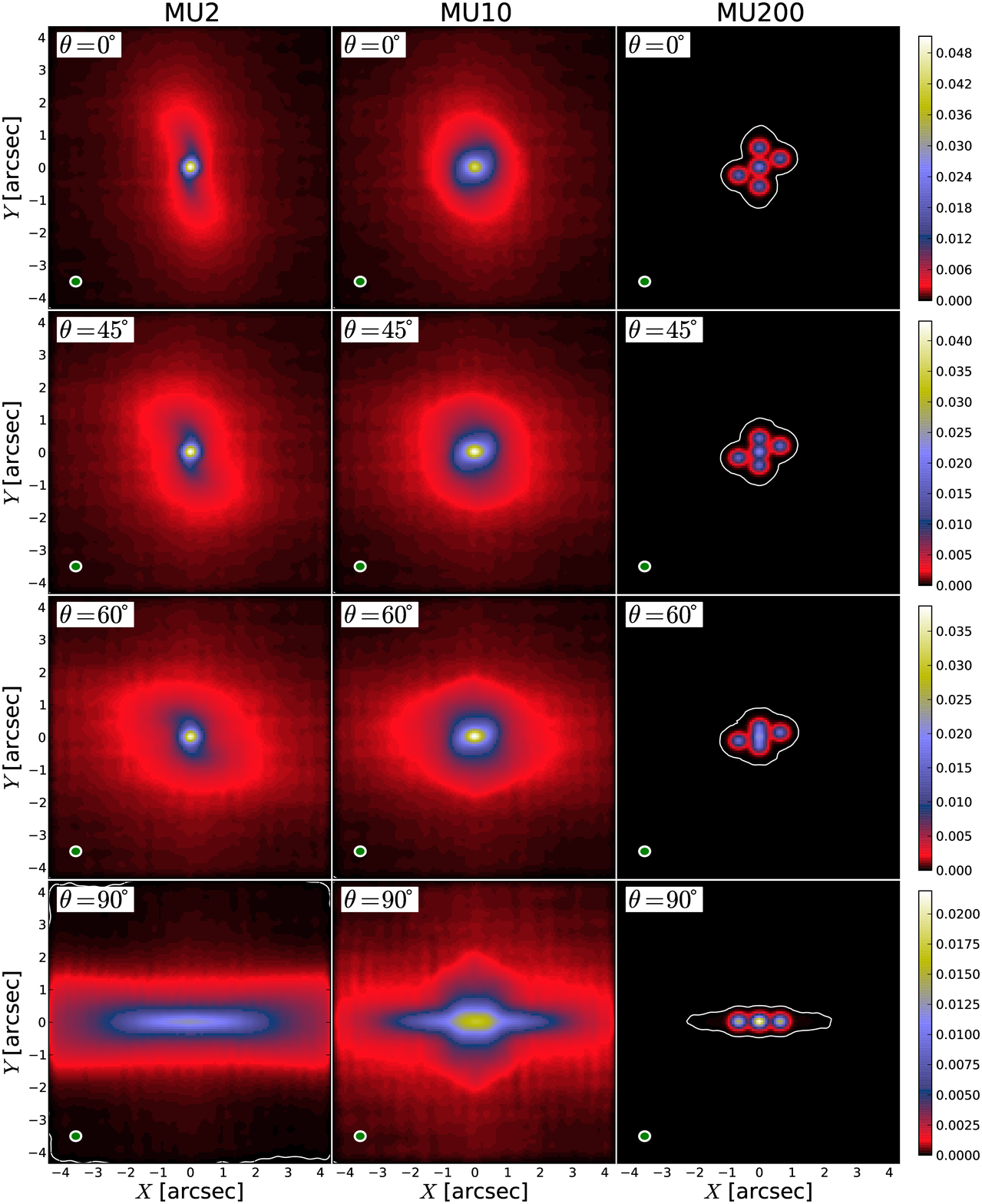}
\caption{Same as Fig.~\ref{c15_b3} but for $B=4$ and $C=15$. Here, $\sigma_S=16.05~\mu{\mathrm{Jy}}$.}
\label{c15_b4}
\end{figure*}
\begin{figure*}
\centering
\includegraphics[scale=0.52]{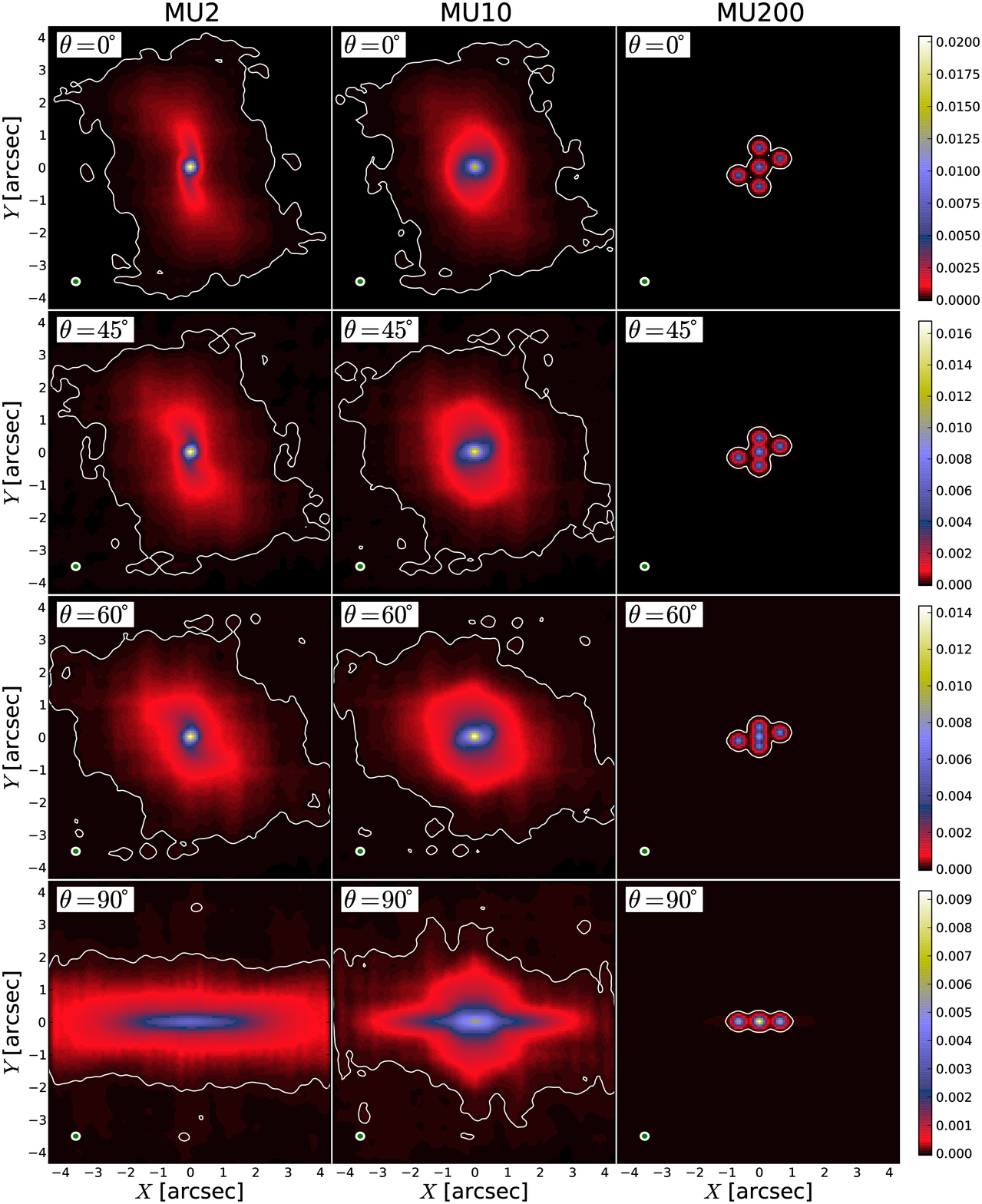}
\caption{Same as Fig.~\ref{c15_b3} but for $B=3$ and $C=20$. Here, $\sigma_S=14.55~\mu{\mathrm{Jy}}$.}
\label{c20_b3}
\end{figure*}
\begin{figure*}
\centering
\includegraphics[scale=0.52]{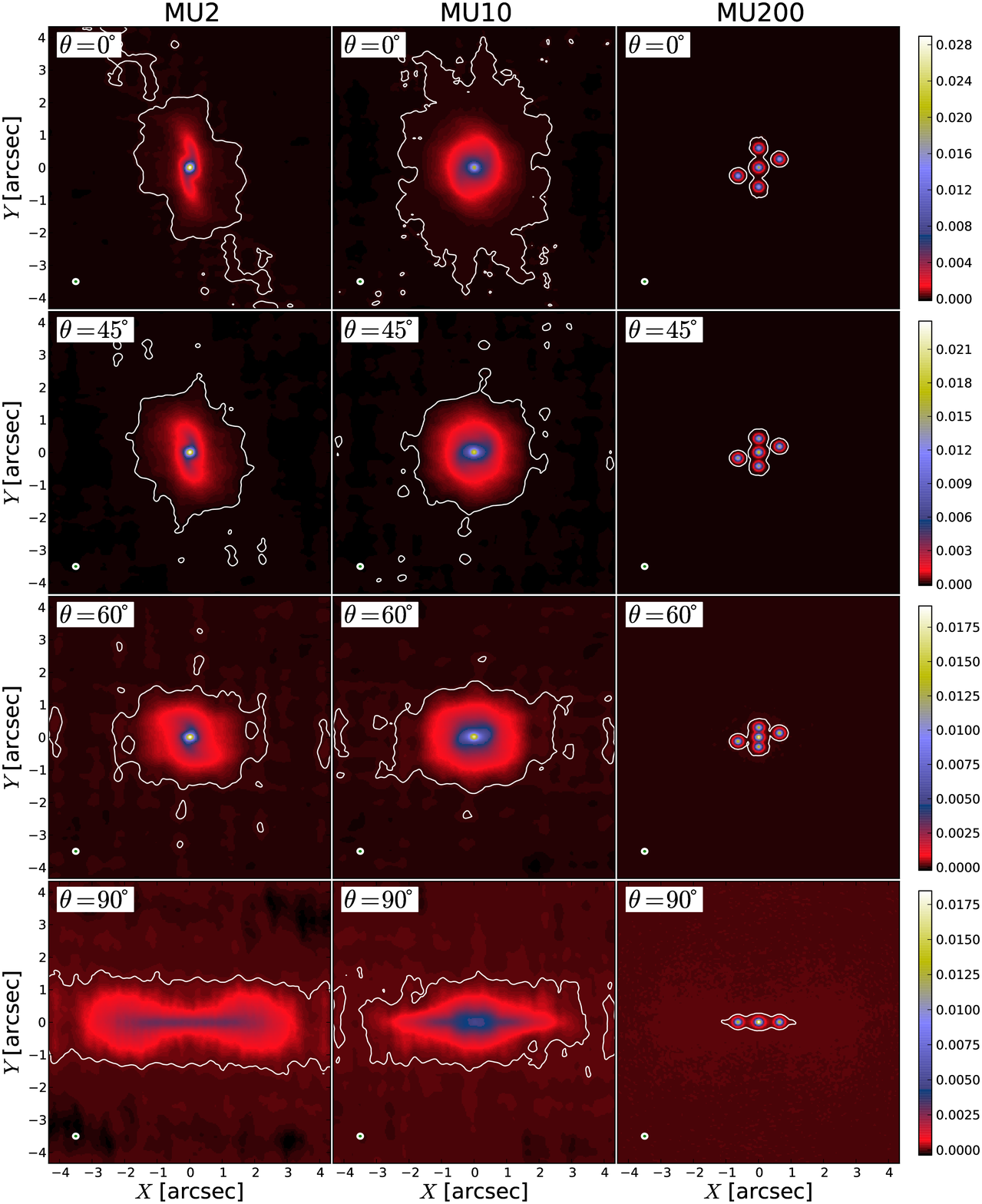}
\caption{Same as Fig.~\ref{c15_b3} but for $B=4$ and $C=20$. Here, $\sigma_S=16.05~\mu{\mathrm{Jy}}$.}
\label{c20_b4}
\end{figure*}

\subsection{Emission maps}

We show, in Fig.~\ref{c15_b3} to~\ref{c20_b4}, the brightness distribution maps output by the ALMA simulator in bands 3 and 4, for configurations 15 and 20. In each figure, four inclination angles are shown, i.e.,  $\theta=0^\circ, 45^\circ, 60^\circ$, and  $90^\circ$. Also shown are the sensitivity limits $3\sigma_S$ in these two bands, given by the ALMA Sensitivity Calculator\footnote{The weather conditions used are that of the first octile, i.e., 0.47mm of precipitable water vapor above the instrument.}. The most extended configuration and the largest frequency best probe the fragmentation scales ($C=20$, $B=4$ in Fig. \ref{c20_b4}). However, as we discussed above, a significant amount of the flux emanating from the extended emission is lost in configuration 20. On the other hand, the fragmentation scale is barely resolved with $C = 15$ and $B = 3$. A better compromise between high resolution imaging and flux recovering is found using either $C = 15$ and $B = 4$ or $C = 20$ and $B = 3$. Overall, there is a clear distinction between the two magnetized models (MU2 and MU10) and the quasi-hydro MU200 model. As expected theoretically, as soon as a magnetic field is taken into account, with field strengths in the range of what is actually {\it{observed}} \citep[e.g.,][]{Falgarone_et_al_2008,Crutcher_et_al_2010,Maury_et_al_2012}, the picture changes dramatically.

Interestingly, many features that are directly probing the physical conditions can be observed with the dust emission already (see Table \ref{table:summary}). In the  MU200 model, the fragmentation is resolved for  inclination angles $\theta<60^\circ$, and only the compact emission of the disk is observed in the edge-on view. In contrast to that,  the dust emission is much more extended in the magnetized MU2 and MU10 models, and corresponds to density features that are typical of magnetized collapse: the pseudo-disk and the outflow. In both cases, the most powerful emission comes from the central FHSC (point-like source,yellow area). In the MU10 model, the disk emission is observed (blue regions) for  inclination angles $\theta<60^\circ$, but it remains relatively small in comparison to the pseudo-disk one. The pseudo-disk emission features are very similar in the MU2 and MU10 models for inclination angles $\theta<60^\circ$. In the edge-on view ($\theta=90^\circ$), the emission from the pseudo-disk is clearly observed in the MU2 and MU10 models, and is much more extended than that of the disk in the MU200 model. The pattern of the pseudo-disk observed with  $C = 20$ and $B = 4$ in the MU2 model may lead to confusion since it has the shape of a flared disk. Additional line emission observations are then needed to distinguish between a disk and a pseudo-disk in that case. Last but not least, the outflow is also observed in the MU10 model in Fig.~\ref{c15_b3} to~\ref{c20_b3}, with an extent of $\sim 4''$ along the polar axis, corresponding to $\sim 300$ AU. This is not surprising since the outflow tends to transport more mass in the case of a lower magnetization \citep[][ and see Fig. 2 in \citealp{Commercon_et_al_2012}]{Hennebelle_Fromang_2008}.

We investigate thoroughly in appendix \ref{sec:noise} the effects of noise (thermal and atmospheric phase) and pointing errors on the synthetic observations. The main conclusions draw from the noise-free case remain unchanged, except that more flux is lost in noisy observations. We refer interested readers to appendix \ref{sec:noise}.

\section{Discussion}

One limitation of this work is that we used a unique dust opacity model from \cite{Semenov_et_al_2003A&A}, in which dust grains can grow, so that we do not predict the variation of the dust emission with different dust grain properties. The dust properties (size, composition, and morphology) within dense cores are still very uncertain, but there is nevertheless theoretical and observational evidence of dust grain growth within dense cores \citep[e.g.,][]{Ormel_et_al_2009,Steinacker_et_al_2010}. \cite{Ormel_et_al_2011} compared the dust grain opacities they got from the various grain models of \cite{Ormel_et_al_2009} with the \cite{Ossenkopf_Henning_1994} opacities model and found that the opacity,  in the submillimeter wavelength range, only varies by a factor of a few between all the different models. We also checked that the \cite{Semenov_et_al_2003A&A} opacities we used are similar to the from the \cite{Ossenkopf_Henning_1994} ones by factors close 1, so that the choice in the opacity will not change the picture.  In addition, we show in Paper I (Fig. 4) that the radius at which the optical depth equals unity does not depend strongly on the opacity for (sub)millimeter wavelengths. We thus speculate that our predictions are relatively robust given the small variations in optical depth and in opacity in the (sub)millimeter wavelength range. 

The second main limitation comes from our idealized initial and boundary conditions, which do not account for the dense core environment (e.g., eventual mass accretion on the core while it collapses) and initial turbulence, whereas the latter can potentially modify the magnetic braking and thus the fragmentation properties during the collapse \citep[e.g., for higher mass dense cores, ][]{Seifried_et_al_2012}. Turbulence, however, is observed to be sub- to trans-sonic in low mass dense cores \citep{Goodman_et_al_1998,Andre_et_al_2007}, so that it will not change dramatically the outcome of the magnetized dense core collapse. In addition, magnetic fields dominate the dynamics and inhibit the fragmentation even more when combined with radiative transfer because of the energy released from the accretion shock at the FHSC border \citep{Commercon_et_al_2010,Commercon_et_al_2011c}. We thus conclude that the three models presented here are representative of the variety of FHSCs that can be formed with various initial conditions. 

Finally, all the analysis and the ALMA synthetic map calculations have been done for objects placed at a distance of 150 pc. The best compromise between flux loss and angular resolution may thus be altered for objects at significantly different distances, but the method remains unchanged.

\section{Summary and perspectives}

We present synthetic dust emission maps of FHSCs as they will be observed by the ALMA interferometer, using state-of-the-art radiation-magneto-hydrodynamic models of collapsing dense cores with different levels of magnetic intensity. We post-process the RMHD calculations performed with the \ttfamily{RAMSES }\rm code using the 3D radiative transfer code \ttfamily{RADMC-3D }\rm to produce dust emission maps. The synthetic observations are then computed using the ALMA simulator within the \ttfamily{GILDAS }\rm software package.

We show that ALMA will shed light on the fragmentation process during star formation and will help in discriminating not only between the different physical conditions, but also between the different models of star formation and fragmentation. The forthcoming facility will yield interferometric observations of FHSC candidates (selected using SED data, see Paper I) in nearby star-forming regions with sufficient angular resolution to probe the fragmentation scale. We also show that the intensity of the magnetic field in this early phase of protostellar collapse will also be assessed, since we do see clear morphological differences between the magnetized and non-magnetized models, i.e., pseudo-disk and outflow in the former case versus disk and fragmentation features in the latter case. We stress that due to the unprecedented sensitivity and coverage of the instrument, this will be achievable in a very short time (18 minutes), so that many FHSC candidates may be observed in a single observing run.

Reaching good enough angular resolution to probe the fragmentation scales and limiting the loss of large-scale emission requires a compromise, which is achieved using frequency bands 3 and 4 and the relatively extended configurations 15 and 20.  We also investigate the effect of noise on the interferometric observations and show that in typical conditions (see Fig. \ref{noisy_map}), ALMA will still be able to reveal fragmentation. We note that the effect of atmospheric phase noise can be efficiently reduced using water vapor radiometers on the ALMA antennas.

We do not discuss the impact of the ALMA Compact Array (ACA) on our simulated observations. Its purpose is to provide measurements for the large-scale emission and improve the wide-field imaging capabilities of ALMA~\citep{alma_memo_398}, but it is not meant to resolve fragmented molecular cores. In our models, this requires a 0.5" angular resolution (see e.g., the case $B=3$ and $C=15$ on Figs.~\ref{uvdistrib} and \ref{c15_b3}), but the longest baselines accessible to ACA are $\sim 100~\mathrm{m}$, so that the best angular resolution available, at the high-frequency end of band 10 (950 GHz), is 0.8". However, ACA could help in recovering some of the lost flux from the disk, pseudo-disk and outflow.

Our work is currently limited to the dust continuum emission, which cannot yet provide robust means to discriminate between FHSC and second hydrostatic cores (SHSC) and to conclude on the nature of VeLLOs. Further work including molecular line emission calculations is thus warranted to better probe the physical conditions (density, temperature, etc...) in observed collapsing cores, for instance to disentangle between the disk and the pseudo-disk, which should harbour different line profiles (rotation dominated versus infall dominated). Line emission predictions are thus the obvious next step towards FHSCs characterization.

\begin{acknowledgements}
We thank the anonymous referee for his/her comments. F.L. wishes to thank J\'er\^ome Pety, Alwyn Wootten and Ian Heywood for their collaboration in including the final ALMA configurations in the {\tt GILDAS} simulator. The research of B.C. is supported by the postdoctoral fellowships from the CNES and by the french ANR Retour Postdoc program.
\end{acknowledgements}
\bibliographystyle{aa}
\bibliography{biblio1}
\begin{appendix}

\section{Noisy observations}
\label{sec:noise}

The simulations presented in the main body of the paper were performed in the unrealistic case of noiseless observations. To truly assess ALMA's ability to resolve fragmented cores, an extension of this study to noisy observations is required, taking into account the different causes of noise : pointing errors, thermal noise and phase noise. 

We start with an error-free simulation of the MU200 model viewed face-on ($\theta=0^{\circ}$), in band $B=4$, with configuration $C=20$. All parameters are identical to those of the general set of simulations, but we add zero-spacing with a single-dish observation, so that in this case all of the flux is recovered (0.1090 Jy) instead of 96\% of it (0.1046 Jy) with ALMA only.

The following errors, whose results are summarized in Table~\ref{table:noise-results}, may then be added, separately or simultaneously, to the simulated observations :
\\
$\blacktriangleright$ 0.6" random pointing errors $\sigma_\alpha$, which correspond to $\sim 15\%$ of the primary beam width at 144 GHz. They may be applied to the ALMA antennas alone or to both ALMA and single-dish measurements. 
\\
$\blacktriangleright$ Thermal noise $\sigma_T$. In band 4, the ALMA Sensitivity Calculator suggests using the 6$^\mathrm{th}$ octile for the atmospheric conditions, which corresponds to 2.75 mm of precipitable water vapor and a zenith opacity $\tau=0.057$. The band 4 receiver noise is set to $T_{144}=40~\mathrm{K}$ following~\cite{asayama_et_al_2008}. Regarding the single-dish measurements, we use a system temperature $T_\mathrm{sys}=100~\mathrm{K}$. The simulator allows for setting none, either or both ALMA and single-dish thermal noises.
\\
$\blacktriangleright$ Atmospheric phase noise $\sigma_\phi$. The ALMA simulator allows for a turbulent atmospheric screen to pass over the interferometer, distorting the waveplanes and causing phase errors. This 2D phase screen is characterized by a second-order structure function that is the combination of three power-laws in three spatial ranges, scaled so that the rms phase difference for a 300-m baseline takes a specific value. In our case, we chose this value to be either 30$^{\circ}$ or 45$^{\circ}$~\citep{alma_memo_398}. Situated at an altitude $z=1000~\mathrm{m}$, the screen passes over the instrument at a windspeed $w=10~\mathrm{m.s^{-1}}$. Calibration, which is done every 26 seconds using a calibrator 2 degrees away from the source perpendicularly to the wind direction, may include the use of water vapor radiometers (WVR), which directly measure the amount of precipitable water vapor along the line of sight in the atmosphere above each antenna.

\begin{figure}
\centering
\includegraphics[scale=0.45]{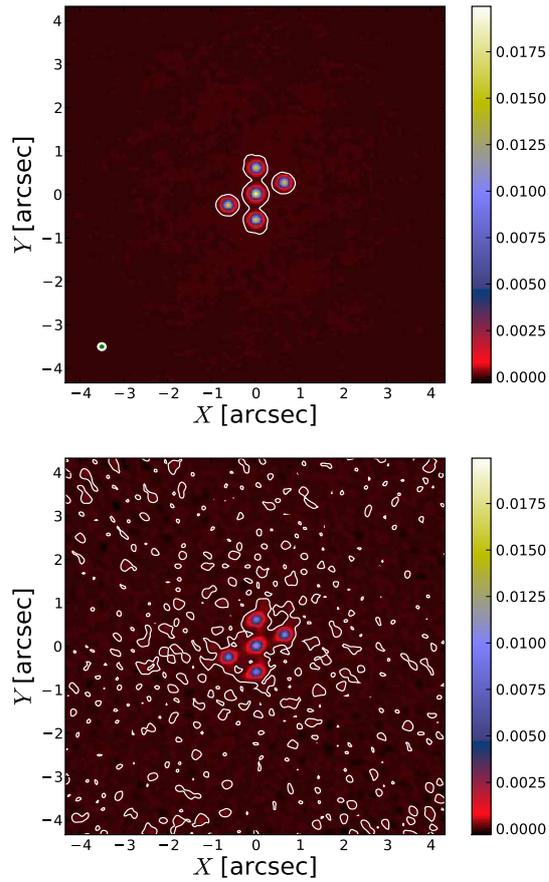}
\caption{Brightness distribution maps, in Jy/beam, for two cases considered in section \ref{sec:noise}. The top plot shows the case of an error-free observation at 144 GHz with ALMA in configuration $C=20$, combined with single-dish. The bottom plot shows the case of an observation with the same combination of instruments, but with 0.6" pointing errors on all antennas, thermal noise as described in the text, and uncorrected atmospheric phase noise with a 45$^{\circ}$ rms phase difference on 300-m baselines. Contours on both plots correspond to the $3\sigma_S$ sensitivity limit given by the ALMA Sensitivity Calculator, where $\sigma_S=17.41~\mu{\mathrm{Jy}}$ for the typical atmospheric conditions in band $B=4$.}
\label{noisy_map}
\end{figure}

\begin{table}
\caption{Effect of the different noise types. The error-free simulation is that of a combined $B=4$ observation of the face-on MU200 model with ALMA in configuration $C=20$ and with a single-dish. Pointing errors ($\sigma_\alpha$) and thermal noise ($\sigma_T$) may be applied to neither instrument ("{\tiny No}"), to the interferometer only ("{\tiny ALMA}"), or to both interferometer and single-dish ("{\tiny Both}"). Phase noise is specified via the rms atmospheric phase on a 300-m baseline, which can take the values 30$^{\circ}$ or 45$^{\circ}$, and can be corrected via water-vapor radiometers ("{\tiny WVR}"). Shown are the fluxes $S$ of the output maps and the median fidelities $F_\mathrm{0.3}$, $F_\mathrm{1}$, $F_\mathrm{3}$ and $F_\mathrm{10}$, on pixels whose intensities in the model image are higher than 0.3\%, 1\%, 3\% and 10\% of the peak, respectively (see section \ref{sec:noise}).}
\label{table:noise-results}
\centering
\begin{tabular}{ccccccccc}
\hline\hline          
$\sigma_\alpha$ & $\sigma_T$ & $\sigma_\phi$ & {\tiny WVR} & $S$ [Jy] & $F_\mathrm{0.3}$ & $F_\mathrm{1}$ & $F_\mathrm{3}$& $F_\mathrm{10}$ \\
\hline
{\tiny No} & {\tiny No} & {\tiny No} & {\tiny No} & {\tiny 0.1090} & {\tiny 107} & {\tiny 208} & {\tiny 279} & {\tiny 418}  \\
{\tiny ALMA} & {\tiny No} & {\tiny No} & {\tiny No} & {\tiny 0.1089} & {\tiny 119} & {\tiny 200} & {\tiny 245} & {\tiny 309}  \\
{\tiny Both} & {\tiny No} & {\tiny No} & {\tiny No} & {\tiny 0.1088} & {\tiny 114} & {\tiny 202} & {\tiny 241} & {\tiny 312}  \\
{\tiny Both} & {\tiny ALMA} & {\tiny No} & {\tiny No} & {\tiny 0.1046} & {\tiny 45} & {\tiny 75} & {\tiny 114} & {\tiny 143}  \\
{\tiny Both} & {\tiny Both} & {\tiny No} & {\tiny No} & {\tiny 0.1035} & {\tiny 44} & {\tiny 75} & {\tiny 112} & {\tiny 141}  \\
{\tiny Both} & {\tiny Both} & {\tiny 30$^{\circ}$} & {\tiny No} & {\tiny 0.0908} & {\tiny 4} & {\tiny 5} & {\tiny 5} & {\tiny 5}  \\
{\tiny Both} & {\tiny Both} & {\tiny 30$^{\circ}$} & {\tiny Yes} & {\tiny 0.0987} & {\tiny 13} & {\tiny 15} & {\tiny 16} & {\tiny 16}  \\
{\tiny Both} & {\tiny Both} & {\tiny 45$^{\circ}$} & {\tiny No} & {\tiny 0.0736} & {\tiny 2} & {\tiny 3} & {\tiny 3} & {\tiny 3}  \\
{\tiny Both} & {\tiny Both} & {\tiny 45$^{\circ}$} & {\tiny Yes} & {\tiny 0.0947} & {\tiny 7} & {\tiny 7} & {\tiny 8} & {\tiny 8}  \\
\hline    
\end{tabular}
\end{table}

Table~\ref{table:noise-results} gives simulation results associated to the various noise situations considered : fluxes in the output (deconvolved) maps, and median fidelities on pixels whose intensities in the model image are higher than 0.3\%, 1\%, 3\% and 10\% of the peak. The fidelity is basically the inverse of the relative error between the output map and the model~\citep{alma_memo_398}, so that the higher the fidelity, the better the reconstruction. What is apparent is that atmospheric phase noise has the strongest impact on the reconstruction process, with a third of the flux being lost in the worst-case scenario, and fidelities dropping to a few (30\%-50\% relative error on the output maps). The use of WVR is a definite plus in this situation, as relative errors then drop to a little over 10\%. However, even in the worst possible situation considered here, and without WVR correction, ALMA still is able to uncover fragmentation, as Fig.~\ref{noisy_map} shows that all five fragments in the model map are well observed above the noise level in these typical conditions.

\end{appendix}
\end{document}